\documentclass[aps,prl,twocolumn,groupedaddress,showpacs]{revtex4}

\usepackage{graphicx}
\usepackage{amsfonts}
\usepackage{amsmath}
\usepackage{amssymb}
\usepackage{bm}

\begin{document}

\title{Robust AC Quantum Spin Hall Effect without Participation of Edge States }
\author{W. Y. Deng$^{1}$}
\author{H. Geng$^{1}$}
\author{Wei Luo$^{1}$}
\author{Wei Chen$^{2}$}
\author{L. Sheng$^{1,3}$}
\email{shengli@nju.edu.cn}
\author{D. N. Sheng$^4$}
\author{D. Y. Xing$^{1,3}$}
\affiliation{$^1$ National Laboratory of Solid State Microstructures and Department of Physics, Nanjing University, Nanjing 210093, China\\
$^2$College of Science, Nanjing University of Aeronautics and Astronautics, Nanjing 210016, China\\
 $^3$ Collaborative Innovation Center of Advanced Microstructures,
Nanjing University, Nanjing 210093, China\\
$^4$ Department of Physics and Astronomy, California State
University, Northridge, California 91330, USA}
\date{\today }

\begin{abstract}
The quantum spin Hall (QSH) effect
in the DC regime, which has been intensively researched, relies on
the existence of symmetry-protected edge states. Here, we demonstrate
that a QSH system behaves quite differently in response to 
an applied AC electric field, and put forward the 
idea of AC QSH effect.  The AC QSH effect can
occur in the bulk without involving the fragile edge states,
hence being robust against time-reversal symmetry breaking and disorder.
It lays a more solid foundation for practical applications
of the two-dimensional topological insulators, in the 
emerging field of AC spintronics.

\end{abstract}

\maketitle

In recent years, there has been a great surge of research interest
in topological insulators (TIs)~\cite{qshe1,qshe2,qshe3,3dti1,3dti2,3dti3,3dti4,3dti5,
rev3DTI1,rev3DTI2,rev3DTI3}. Two-dimensional TIs,
also called quantum spin Hall (QSH) systems~\cite{qshe1,qshe2,qshe3},
are featured with
a bulk band gap around the Fermi level and gapless helical edge
states traversing the band gap.
The topological distinction between the QSH materials and ordinary
insulators can be
described by an unconventional topological invariant, i.e., the
Z$_2$ index~\cite{Z2} or spin Chern number~\cite{spinch1,spinch2,spinch3}.
While the existence of edge states in the QSH systems
is attributable to the nontrivial bulk band topology,
the gapless nature of the edge states is protected
by the time-reversal (TR) symmetry~\cite{qshe1,qshe2,qshe3}.
The QSH effect provides a purely electrical
means to generate dissipationless and noiseless spin current and 
spin accumulations, and so is promising
for applications in high-precision low-power spintronic devices.

Most existing research works about the QSH effect 
focus on the steady states in the DC regime. In a steady state in the DC regime,
only electrons at the Fermi level can contribute to the electronic transport,
as indicated clearly by the Landauer-B\"{u}ttiker formalism, so that
the DC QSH effect must be carried by the symmetry-protected
gapless edge states that pass through the Fermi level.
In the presence of TR-symmetry-breaking perturbations,
the edge states become gapped~\cite{qshe1,qshe2,qshe3} and can be
localized by any weak disorder due to their one-dimensional nature,
according to the textbook theory of Anderson localization.
As a consequence, the DC QSH effect
is fragile in realistic environments, where such symmetry-breaking perturbations
are often unavoidable. So far, conductance near the predicted quantized value
through  edge channels  has been realized 
only in small samples of
HgTe quantum wells~\cite{HgTe} and InAs/GaSb
bilayers~\cite{RRDu}, with sizes of the order of $1\mu m\times 1\mu m$, 
and the conductance is strongly suppressed for larger samples.  
This is in contrast to the conventional quantum Hall effect, 
for which precisely integer-quantized Hall
conductivity has been observed
in a large variety of materials on the macroscopic scale.
The instability of the DC QSH effect remains to be a serious
technical barrier to its practical applications.


\begin{figure}
\includegraphics[width=2.6in]{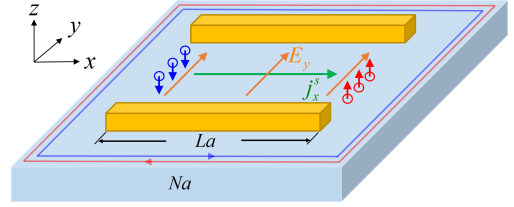}
\caption{Proposed experimental setup, and illustration of the AC QSH effect
discovered in this Letter.
Two metal bars are deposited on the surface of a QSH sample, keeping
enough distances from the sample edges.
By applying an AC bias voltage to the metal bars, an AC electric field
$E_{y}(t)=E_{0}\cos\omega t$ is
created  along the $y$ direction in the region between the bars (biased region).
In response, an AC spin current is generated in the $x$
direction within the biased region, which in turn results in opposite measurable spin
accumulations, oscillating with time, near the right and left 
edges of the biased region. The whole
transport process does not involve the edge states.
}\label{Fig1}
\end{figure}

\begin{figure*}
\includegraphics[width=5.2in]{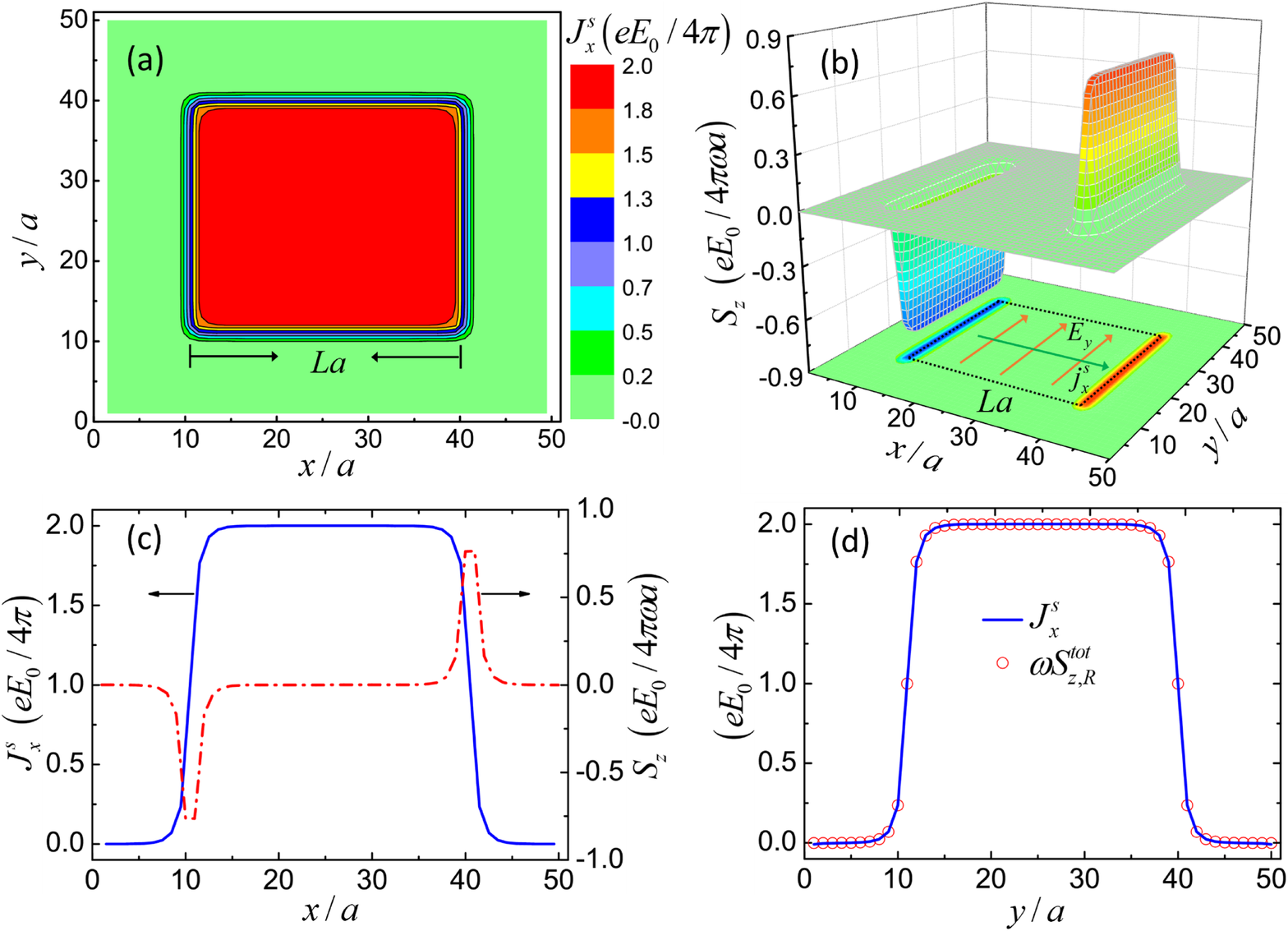}
\caption{(a) Profile of spin current density $J_{x}^{s}(x,y)$ in the whole sample. (b) Density of spin accumulation $S_{z}(x,y)$
 in the whole sample. (c) Spin current density and density of spin accumulation
along the mirror line $y=\frac{Na}{2}$ as functions of $x/a$. (d)
Spin accumulation rate $\omega S_{z,R}^{tot}(y)$ and spin current density
$J_{x}^{s}(\frac{Na}{2},y)$ as functions of $y$. The sizes of the sample
and biased region are set to $N=50$ and $L = 30$, respectively. The other parameters
are taken to be $\hbar \omega = 0.05$, and the Fermi energy $E_{\rm{F}}=0$,
in the middle of the band gap.
}\label{Fig2}
\end{figure*}

Recently, AC spin current 
in the G-Hz frequency range generated through spin pumping 
has been detected experimentally, which stimulates the emerging field
of AC spintronics~\cite{ACspinC1}. In an AC electric field, 
a QSH system will always be in a transient state,
which may display quite different 
characteristics from the steady state in the 
DC regime. In this Letter, we demonstrate that the AC QSH effect is 
much more stable than its DC counterpart.
It can occur in the bulk
of a QSH material without involving the fragile edge states, if the driving
AC electric field is applied only to an inner region of the
sample (as illustrated in Fig.\ 1). 
We perform exact finite-size
numerical calculations, adopting the BHZ model~\cite{BHZ} for
the QSH effect, based on the Kubo linear-response theory.
We show that when electron spin is conserved, a quantized transverse AC
spin current is generated within the region of the
electric field (biased region), and
measurable spin accumulations, oscillating with time, occur near the transverse edges
of the biased region, as required by the spin conservation
law. Moreover,  even when magnetic disorder
is present, breaking both spin conservation and TR symmetry,
the spin accumulations remain robust, although the spin current
becomes ill-defined. The AC QSH effect
lays a more solid foundation for practical applications
of the two-dimensional TIs in spintronics, immune to
symmetry breaking and disorder.

To study the AC QSH effect numerically,
we adopt the lattice version of the BHZ model~\cite{BHZ}, which can be constructed
on a square lattice from the continuum Hamiltonian~\cite{BHZ} through the replacements ${k_{x}} \to {a^{ - 1}}\sin \left( {{k_{x}}a} \right)$, ${k_{y}} \to {a^{ - 1}}\sin \left( {{k_{y}}a} \right)$ and ${k^2} \to 2{a^{ - 2}}\left[ {2 - \cos \left( {{k_x}a} \right) - \cos \left( {{k_y}a} \right)} \right]$ with $a$
as the lattice constant. After a reverse
Fourier transformation, the Hamiltonian in real space is obtained as
\begin{eqnarray}
{H_0} &=&\sum\limits_{\bf r} {c_{\bf r}^\dag {\varepsilon _0}{c_{\bf r}}}  + \sum\limits_{\bf r} {\left( {c_{\bf r}^\dag {t_x}{c_{{\bf r} + {a\hat{x}}}} + {\rm{h}}{\rm{.c}}{\rm{.}}} \right)}\nonumber \\
 &+& \sum\limits_{\bf r} {\left( {c_{\bf r}^\dag {t_y}{c_{{\bf r} + {a\hat{y}}}} + {\rm{h}}{\rm{.c}}{\rm{.}}} \right)}\ ,
\end{eqnarray}
where ${\bf r}= \left( x,y \right)$ are the coordinates
of a lattice site, and $c_{\bf r}  = \left( {c_{{\bf r}s\uparrow} ,c_{{\bf r}p\uparrow} ,c_{{\bf r}s\downarrow} ,c_{{\bf r}p\downarrow}}\right)$ are the electron annihilation operators
on site ${\bf r}$ with $\uparrow$ and $\downarrow$ as
spin indices and $s$ and $p$ representing
two orbits. Here, ${\varepsilon _0} =  - 4\left( {D - B{\sigma _z}} \right){a^{ - 2}} - {M_0}{\sigma _z}$, ${t_x} = \left( {D - B{\sigma _z}} \right){a^{ - 2}} - i{\left( {2a} \right)^{ - 1}}A{\sigma _x}{s_z}$, and ${t_y} = \left( {D - B{\sigma _z}} \right){a^{ - 2}} + i{\left( {2a} \right)^{ - 1}}A{\sigma _y}$. ${\mbox{\boldmath{$\sigma$}}}
=(\sigma_{x},\sigma_{y},\sigma_{z})$ and ${\mbox{\boldmath{$\textbf{s}$}}}
=(s_{x},s_{y},s_{z})$ are the Pauli matrices for orbit and spin, respectively, and $A$, $B$, $D$ and ${M_0}$ are the parameters
of the BHZ model~\cite{BHZ}. The topological properties of the
model can be described by the topological invariant, spin Chern number, given by ${C_{s}} = {{\mathop{\rm sgn}} \left( B \right) + {\mathop{\rm sgn}} \left( {{M_0}} \right)}$~\cite{HgTeSpinCh}. Since $C_{s}$ is independent of $D$, without loss of generality, we set $D=0$ in the
following calculations. The Dirac mass $M_{0}$ (assumed to be positive) is
taken as the unit of energy.  The other model parameters are
set to $A/a = 3$, and $B/a^{2} = 1.5$.
For this set of parameters, the BHZ model is in the QSH phase with ${C_{s}} = 2$.

We consider a square sample of size ${N}a$ with open boundary.
To demonstrate the independence of the AC QSH effect on the 
edge states, it is assumed that an AC electric field $E_{y}\left( t \right) = {E_0}\cos \left( {\omega t} \right)$ is applied along the $y$ direction in a smaller
concentric square region of size ${L}a$, as illustrated in Fig.\ 1.
We study first the ideal case, where electron spin is conserved.
The Kubo linear-response theory~\cite{Mahan} is employed to calculate the spin current
density $j_{x}^{s}(x,y; t)$ and density of spin
accumulation $\rho_{s}(x,y; t)$ induced by the applied electric field.
We find that the spin current and spin accumulations are always polarized
in the $z$ direction. Besides, for sufficiently large $La$,
the longitudinal component of the spin current along
the $y$ direction is negligible, essentially because the system
has an insulating gap around the Fermi level.
The spin current response to the driving AC electric field is immediate, and so
synchronized with the electric field, i.e., $j_x^s\left(x,y;t \right) = J_{x}^{s}(x,y)\cos \left( {\omega t} \right)$, where $J_{x}^{s}(x,y)$ is obtained from
the Kubo formula~\cite{Mahan} as
\begin{equation}
J_{x}^{s}(x,y) = \frac{{{\hbar ^2}eE_{0}}}{{{a^2}}}\sum\limits_{m \ne n} {\frac{{{\mathop{\rm Im}\nolimits} \left[ {\left\langle {m\left| {{v_y}} \right|n} \right\rangle \left\langle {n\left| {v_x^s} \right|m} \right\rangle } \right]}}{{{{\left( {{E_m} - {E_n}} \right)}^2} - {{\left( {\hbar \omega } \right)}^2}}}f\left( {{E_m}} \right)}\ .
\end{equation}
Here, the eigenstates $\left| m \right\rangle$ with eigenenergies ${E_m}$ are obtained through exact diagonalization of the Hamiltonian $H_{0}$ of the sample, $f(E)$ is the Fermi distribution function, ${v_y} = \frac{i}{\hbar }\left[ {H_{0},y} \right]$
is the velocity operator in the biased region,
and $v_x^s = \frac{1}{2}\left\{ {{v_x},{s_z}} \right\}$ represents the spin velocity operator  at $(x,y)$.

In Fig.\ \ref{Fig2}(a), we show the profile of the
calculated spin current density $J_{x}^{s}(x,y)$
for a small frequency in the whole sample. We can see that  inside the biased
region,  the ratio $J_{x}^{s}(x,y)/E_{0}\equiv j_{x}^{s}(x,y;t)/E_{y}(t)$, 
being equivalent to the frequency-dependent spin Hall conductivity,
takes an integer-quantized value $2$ in units of the
spin conductivity quantum $\frac{e}{4\pi}$, which is well consistent with the spin Chern number $C_{s}=2$. 
The spin current density $J_{x}^{s}(x,y)$
vanishes quickly crossing the right and left edges of the biased region.
The discontinuities in
$J_{x}^{s}(x,y)$ imply that electron spins
are accumulating or depleting near the edges.

\begin{figure}
\includegraphics[width=2.7in]{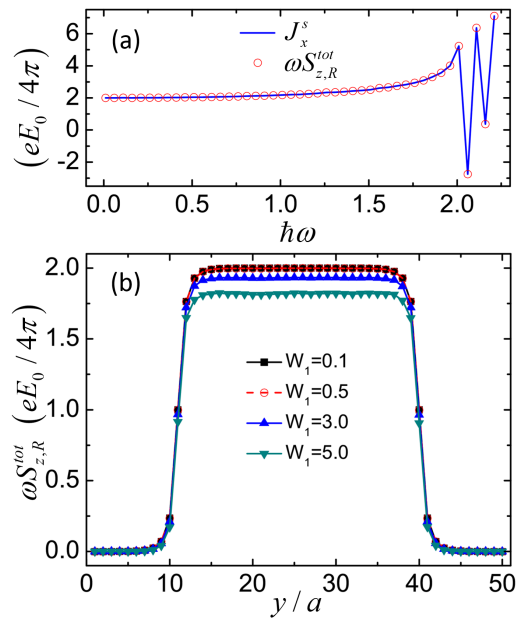}
\caption{(a) Spin accumulation rate $\omega S_{z,R}^{tot}(y)$ and
spin current density $J_{x}^{s}(\frac{Na}{2},y)$ at $y=\frac{Na}{2}$
as functions of frequency. (b) Spin accumulation rate $\omega S_{z,R}^{tot}(y)$
as a function of $y/a$ for some different magnetic disorder strengths
with nonmagnetic disorder strength set to $W_{0}=0$. Each data point is averaged over 300 disorder configurations. The other parameters are the same as in Fig.\ 2.
}\label{Fig3}
\end{figure}

The density of spin accumulation
is related to the spin current density through the
spin conservation law
\begin{equation}\label{conserv}
\frac{{\partial {\rho _s}}}{{\partial t}} + \frac{{dj_x^s}}{{dx}} = 0\ .
\end{equation}
It follows that the density of spin accumulation is an integral over time
of the gradient of the spin current density, and thus lags the
spin current by $\pi/2$. As a
result, $\rho_{s}(x,y; t) = {S_z}(x, y)\sin \left( {\omega t} \right)$,
where ${S_z}(x,y)$ can be obtained from the Kubo formula~\cite{Mahan} as
\begin{eqnarray}\label{Kubo_Sz}
{S_z}(x,y) &=& \frac{{\hbar eE_{0}}}{{\omega {a^2}}}\sum\limits_{m \ne n}
\left( {{E_m} - {E_n}} \right)\nonumber 
\\
&\times&{\frac{{{\mathop{\rm Re}\nolimits} \left[ {\left\langle {m\left| {{v_y}} \right|n} \right\rangle \left\langle {n\left| {{s_z}} \right|m} \right\rangle } \right]}}{{{{\left( {{E_m} - {E_n}} \right)}^2} - {{\left( {\hbar \omega } \right)}^2}}}f\left( {{E_m}} \right)}\ .
\end{eqnarray}
Here, $s_z$ stands for the Pauli matrix for electron spin
at $(x,y)$. In Fig.\ \ref{Fig2}(b), we plot the
calculated profile of the density of
spin accumulation in the whole sample. Indeed, there are
spin accumulations sharply peaked at the right and left edges of the biased
region. In Fig.\ \ref{Fig2}(c), the spin current density
and density of spin accumulation along the mirror line $y=\frac{Na}{2}$
as functions of $x/a$ are plotted.
One can see more clearly that the peaks of the spin accumulations
correspond to the discontinuities in the spin current density.

We wish to point out that the value $\hbar\omega = 0.05$ of the frequency
set in the above calculations
is already very small compared with the band gap $2$. We have also studied 
the cases for even much smaller frequencies, e.g., $\hbar\omega = 0.001,$ 
$0.0001$, and the  transverse
profile of the spin current is almost unchanged.  
This means that the discontinuities in the 
spin current near the edges of the biased 
region will persist in the limit $\omega\rightarrow 0$,  
which is essentially different from the DC regime.
In the DC regime, $\frac{{\partial {\rho _s}}}{{\partial t}}=0$
in a steady state, so the spin current must be continuous 
throughout the sample. In fact, the DC regime has been known to be singular for a long time~\cite{Mahan}.
The vector potential of an exactly static electric field $E_{0}\hat{y}$
may be chosen to be $-E_{0}t\hat{y}$, which is unbounded and can not
be treated within the Kubo linear-response theory. To overcome this difficulty, 
the DC regime is usually regarded as being equivalent to the $\omega\rightarrow 0$ 
limit of the AC regime~\cite{Mahan}. 
This assumption is however not necessarily valid always. The characteristics of the
transient-state spin current and spin accumulations at small frequencies revealed in this work 
is not extensible to the steady state in the DC regime at $\omega=0$.

To further show quantitatively that the calculated results
conform the spin conservation law, we readily
derive from Eq.\ (\ref{conserv}) for the following relations between the
spin current density and density of
spin accumulation
\begin{eqnarray}\label{StotR}
\omega S_{z,R}^{tot}(y)&\equiv& \omega \int_{\frac{{{Na}}}{2}}^{Na}{{S_z(x,y)}dx} =
J_{x}^{s}(\frac{Na}{2},y)\ ,\\
\omega S_{z,L}^{tot}(y)&\equiv& \omega\int_0^{\frac{{{Na}}}{2}}{{S_z(x,y)}dx} =
-J_{x}^{s}(\frac{Na}{2},y)\ .
\end{eqnarray}
Here, we have taken into account the fact that the spin current density essentially
vanishes at the sample edges, i.e., $J_{x}^{s}(0,y)=J_{x}^{s}(Na,y)=0$,
as can be seen from Fig.\ \ref{Fig2}(a). Apparently,
$S_{z,R}^{tot}(y)$ and $S_{z,L}^{tot}(y)$ are
the total spin accumulations per unit width (in the $y$ direction)
around the right and left edges of the biased region, respectively.
The quantities $\omega S_{z,R}^{tot}(y)$ and $\omega S_{z,L}^{tot}(y)$
essentially measure the changing rates of the total spin accumulations.
Since $S_{z,L}^{tot}(y)=-S_{z,R}^{tot}(y)$, we will consider $S_{z,R}^{tot}(y)$
only. In Fig.\ \ref{Fig2}(d), both $\omega S_{z,R}^{tot}(y)$ and
$J_{x}^{s}(\frac{Na}{2},y)$ are plotted as functions of $y$.
The two curves coincide with each other.
In Fig.\ \ref{Fig3}(a), the spin accumulation rate $\omega S_{z,R}^{tot}(y)$ and
spin current density $J_{x}^{s}(\frac{Na}{2},y)$
at $y=\frac{Na}{2}$ are displayed as functions of frequency, which again coincide with each other. These results are in good agreement with Eq.\ (\ref{StotR}). From Fig.\ \ref{Fig3}(a),
we also see that at low frequencies, $J_{x}^{s}(\frac{Na}{2},\frac{Na}{2})$
as well as  $\omega S_{z,R}^{tot}(\frac{Na}{2})$,
divided by $E_{0}$, are
quantized to $2(\frac{e}{4\pi})$, but obvious deviations from the
quantized value are observed at large frequencies. Oscillations occur
when the frequency $\hbar\omega$ exceeds the band gap $2$.

\begin{figure}
\includegraphics[width=2.8in]{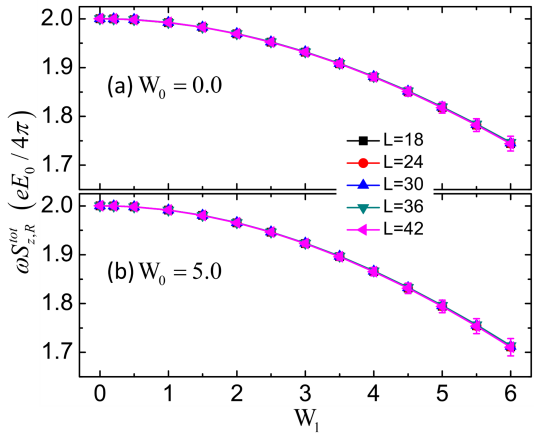}
\caption{Spin accumulation rate $\omega S_{z,R}^{tot}(\frac{Na}{2})$
as a function of magnetic disorder strength, for
size of the biased region ranging from $L=18$ to $42$.
The size of sample is set to $N=L+2\times 10$. The nonmagnetic disorder strength is chosen to be (a) ${W_{0}=0}$ and (b) ${W_0=5}$. Each data point is averaged over 200-500 disorder configurations.
}\label{Fig4}
\end{figure}

In order to demonstrate the robustness of
the AC QSH effect, we introduce
a generalized version of the Anderson disorder into the system. The total Hamiltonian of the system is then given by $H = {H_0} + U$ with
\begin{equation}\label{5}
U = \sum\limits_{\bf r} {c_{\bf r} ^\dag \left( {w_{\bf r} ^{(0)} + w_{\bf r}^{(1)}\textbf{s} \cdot {\textbf{m}_{\bf r} }} \right){c_{\bf r} }}\ ,
\end{equation}
where ${w_{\bf r} ^{(0)}}$ represents on-site nonmagnetic 
disorder, uniformly distributed in the interval ${w_{\bf r} ^{(0)}}\in \left[ { - \frac{W_0}{2},
\frac{W_0}{2}} \right]$, ${w_{\bf r} ^{(1)}}$ stands for on-site magnetic disorder, distributed between ${w_{\bf r} ^{(1)}}\in\left[ { - \frac{W_1}{2},\frac{W_1}{2}} \right]$, and ${\textbf{m}_{\bf r}}$ is a randomly oriented unit vector. Apparently, the presence of the magnetic disorder destroys both the spin conservation and TR symmetry of the system.
It is well-known that the spin current density is ill-defined in the absence
of spin conservation. Nonetheless, from the above discussions, we know that
the spin accumulation rate $\omega S_{z,R}^{tot}(y)$ is equivalent to
the spin current density, which remains well-defined. The Kubo formula
Eq.\ (\ref{Kubo_Sz}) is general and still valid in the presence of disorder.
In Fig.\ \ref{Fig3}(b),
the spin accumulation rate calculated from Eq.\ (\ref{Kubo_Sz})
 is plotted as a function of $y/a$ for several
different strengths of magnetic disorder with nonmagnetic disorder
strength set to
$W_{0}=0$, where each data point is averaged over 300 disorder configurations.
We see that even intermediately strong magnetic
disorder only suppresses the spin accumulations slightly.

We also carry out scaling analysis to study if the AC QSH effect remains
stable for large samples. The calculated $\omega S_{z,R}^{tot}(\frac{Na}{2})$ for
size $La$ of the biased region ranging from $18a$ to $42a$
is displayed as a function of
magnetic disorder strength $W_{1}$. The nonmagnetic disorder strength is
set to $W_{0}=0$ and $5$ in Figs.\ {\ref{Fig4}}(a) and {\ref{Fig4}}(b), respectively. In all calculations, the distance from the biased
region to sample edges is fixed at $10a$. From the figures, we observe that
while intermediately strong disorder suppresses the spin accumulations slightly,
the spin accumulations do not decrease with increasing the sample size. Thus,
we can expect that the AC QSH effect will stay stable for very large samples.

In summary, we have shown that while both the DC and AC QSH effects
originate from the nontrivial bulk band topology of the two-dimensional
TIs, they exhibit quite different characteristics. The DC 
QSH effect must be carried by the symmetry-protected edge states. The AC
QSH effect can occur in the bulk without involving the fragile edge states,
hence being robust against symmetry breaking and disorder. Based upon the AC QSH effect, 
an array of asynchronized electrical spin polarization generators can be built
on a single film of QSH material, which is very beneficial for
designing and fabricating spintronic integrated circuits. Moreover, the low-frequency
AC QSH effect provides a relatively easy way to experimentally investigate
the unconventional topological invariant underlying the QSH systems directly,
compared with previous theoretical proposals, including topological magnetoelectric
effect~\cite{Magneto1,Magneto2} and
topological spin pumping~\cite{TopoPump}.

This work was supported by the State Key Program for Basic Researches of China under
grants numbers 2015CB921202 and 2014CB921103 (L.S.), the National Natural Science Foundation of China under grant numbers 11225420 (L.S.), and a project funded by the PAPD of Jiangsu Higher Education Institutions (L.S. and D.Y.X.). This work was also supported  by the U.S. Department of Energy, Office of 
Basic Energy Sciences under Grant No. DE-FG02-06ER46305 (D.N.S).\\
\\
W.Y.D. and H.G. contributed equally to this work.


\end{document}